\documentclass[aps,physrev,reprint,superscriptaddress,amsmath,amssymb]{revtex4-2}
\usepackage{graphicx}
\usepackage{dcolumn}
\usepackage{hyperref}
\usepackage[mathlines]{lineno}
\usepackage{bbold}
\usepackage{tabularx}
\usepackage{cleveref}
\usepackage{physics}
\usepackage{siunitx}
\usepackage{multirow}
\usepackage{subcaption}
\usepackage{caption}
\usepackage{booktabs}

\begin{document}

\title{Canonical Quantization of a Memristive Leaky Integrate-and-Fire Neuron Circuit}

\author{Dean Brand}
\email{dean.brand@nithecs.ac.za}
\affiliation{Department of Physics, Stellenbosch University, Stellenbosch, 7604, South Africa}
\affiliation{National Institute for Theoretical and Computational Sciences (NITheCS), Stellenbosch, 7604, South Africa}
\author{Domenica Dibenedetto}
\email{domenica.dibenedetto@maastrichtuniversity.nl}
\affiliation{Department of Advanced Computing Sciences, Maastricht University, Maastricht, Netherlands}
\author{Francesco Petruccione}
\email{francesco.petruccione@nithecs.ac.za}
\affiliation{National Institute for Theoretical and Computational Sciences (NITheCS), Stellenbosch, 7604, South Africa}
\affiliation{School of Data Science and Computational Thinking, Stellenbosch University, Stellenbosch, 7604, South Africa}

\date{\today}

\begin{abstract}
    We present a theoretical framework for a quantized memristive Leaky Integrate-and-Fire (LIF) neuron, uniting principles from neuromorphic engineering and open quantum systems. Starting from a classical memristive LIF circuit, we apply canonical quantization techniques to derive a quantum model grounded in circuit quantum electrodynamics. Numerical simulations demonstrate key dynamical features of the quantized memristor and LIF neuron in the weak-coupling and adiabatic regime, including memory effects and spiking behavior. Applications of this model to a sound localization benchmark show that it outperforms a phenomenological quantum LIF model as well as a classical LIF. This work establishes a foundational model for quantum neuromorphic computing, offering a pathway towards biologically inspired quantum spiking neural networks and new paradigms in quantum machine learning.
\end{abstract}

\maketitle

\section{Introduction}
\label{sec:Introduction}

Modern computing is reaching fundamental physical and architectural limits. On the physical side, continued transistor miniaturization is approaching atomic scales, where quantum tunneling induces intolerable error rates that threaten the reliability of classical processors \cite{waldrop_chips_2016}. Architecturally, the von Neumann bottleneck imposes a hard constraint on performance and energy efficiency by requiring continuous data shuttling between memory and processor \cite{zou_breaking_2021}. In contrast, biological brains seamlessly integrate memory and computation within the same substrate, achieving high parallelism and unmatched energy efficiency \cite{benjamin_neurogrid_2014, merolla_million_2014, davies_loihi_2018}.

These two limitations have catalyzed the development of alternative computational paradigms. Quantum computing addresses the physical scaling limit by harnessing superposition and entanglement to perform certain tasks with exponential speed-up, thus circumventing the need for denser circuitry. Neuromorphic computing, on the other hand, draws inspiration from biological neural systems to overcome architectural inefficiencies. By co-locating memory and processing and leveraging spiking neurons and plastic synapses, neuromorphic architectures offer low-power, massively parallel information processing. One key enabling component is the memristor, a resistor with memory, which supports localized learning and efficient connectivity \cite{markovic_physics_2020}.

However, quantum and neuromorphic computing typically pursue separate trajectories: the former excels in computational power and algorithmic advantage for specific problems where it scales better than classical computing, while the latter offers efficient structure and sparsity akin to biological neurons. Both have their strengths but lack in certain areas as well, particularly each other's. This complementarity has motivated a convergence into quantum neuromorphic computing, a nascent but promising field aiming to unify quantum dynamics with brain-inspired architectures \cite{markovic_quantum_2020, sato_study_2004, pehle_neuromorphic_2022}. Though the landscape is vast, encompassing hardware implementations, computational primitives, and learning models, it remains largely unexplored.

Recent contributions span quantum reservoir computing \cite{katumba_neuromorphic_2018, ghosh_quantum_2019, ghosh_quantum_2021}, quantum memristors \cite{pfeiffer_quantum_2016, sanz_invited_2018, guo_y_-m_quantum_2021}, and various proposals for quantum neurons \cite{cao_quantum_2017, tacchino_artificial_2019, kristensen_artificial_2021}. In particular, a gap remains in developing biologically inspired quantum neuron models that balance computational efficiency with biological realism. In classical neuroscience and AI, neuron models range from the highly efficient yet abstract artificial neural networks to the biophysically grounded Hodgkin-Huxley model \cite{hodgkin_quantitative_1952}.

While both extremes have seen some level of quantization, an important intermediate model, the Leaky Integrate-and-Fire (LIF) neuron, has not yet been formally quantized. The LIF model provides a biologically plausible, yet computationally lightweight, spiking framework well-suited to time-dependent tasks and neuromorphic hardware \cite{yamazaki_spiking_2022, wang_supervised_2020}. Although a phenomenological quantum analogue of the LIF neuron has recently been proposed \cite{brand_quantum_2024}, a formal quantum formulation remains lacking. Developing such a model is critical for establishing a theoretical foundation upon which future quantum neuromorphic architectures can be systematically built. Such foundational descriptions will allow for more resource-efficient and computationally expressive models and algorithms to be developed, while being rooted in the fundamental behaviors of quantum systems exhibiting neuromorphic behaviors. The most notable neuromorphic systems to emulate are those of the Hodgkin-Huxley neuron \cite{gonzalez-raya_quantized_2019,gonzalez-raya_quantized_2020} and the LIF neuron, which is presented here.

In this work, we present a direct quantization of a memristive LIF neuron circuit, introducing a quantum memristor model based on explicit bath mode coupling and an open quantum systems framework. Inspired by prior quantizations of biologically grounded models such as the Hodgkin-Huxley neuron \cite{gonzalez-raya_quantized_2019, gonzalez-raya_quantized_2020}, our approach yields a fully quantum spiking neuron with intrinsic memory encoded in its circuit dynamics. We derive a microscopic Hamiltonian for an LC (inductor-capacitor) oscillator --- representing the neuronal membrane --- interacting with a state-dependent transmission line that captures memristive behavior. In the adiabatic and weak-coupling limit, we show analytically how this model recovers the classical LIF dynamics with a memristive leak term.

The rest of this article is structured as follows. We recap the classical formulation of the neuromorphic circuitry of the LIF neuron, and its memristive formulation. Following this we begin the quantization process, resulting in a fully quantized memristive LIF circuit model. Next we verify the derivations with numerical simulations of the models presented and how they match the characteristic features of the classical counterparts. This is demonstrated with an analysis and comparison of the model on a sound localization task along with the classical and phenomenological quantum LIF models. Finally we summarize our findings and discuss the outlook of how this model serves as a foundation for further exploration and development of quantum neuromorphic computing.

\section{Classical Models}
\label{sec:Classical_Models}

We begin by introducing the classical models that underpin our quantum neuromorphic framework: the LIF neuron and the memristor. We show how these two components can be naturally combined to form a compact, biologically inspired circuit that serves as the classical counterpart to our quantum model.

\subsection{Leaky Integrate-and-Fire Neuron}
\label{sec:Leaky_Integrate_and_Fire_Neuron}

The LIF neuron is a foundational model in computational neuroscience and neuromorphic engineering. It captures the essential behavior of biological neurons using a simple electrical circuit. In its standard form, the LIF neuron is modeled as a resistor-capacitor (RC) low-pass filter, where: $C_m$ is the membrane capacitance, $R$  is the leak resistance, $V(t)$ is the membrane potential, and $I_\mathrm{in}(t)$ is the input current. The governing equation is:
\begin{equation}
    \label{eq:rc_diff}
    C_m \dv{V(t)}{t} = -\frac{V(t)}{R} + I_\mathrm{in}(t).
\end{equation}
Here, the capacitor integrates incoming current over time, while the resistor causes the potential to ``leak" back toward rest. When $V(t)$ reaches a threshold $V_\mathrm{th}$, the neuron emits a spike and the potential is reset. Despite its simplicity, this model captures core spiking behavior and serves as the basis for many hardware implementations.

\subsection{Memristor}
\label{sec:Memristor}

To introduce adaptive behavior and memory, we use the memristor, a nonlinear, history-dependent resistor proposed by Chua in 1971 \cite{chua_memristor-missing_1971, chua_memristive_1976}. It completes the set of fundamental passive circuit elements (resistor, capacitor, inductor, memristor) and is especially relevant for neuromorphic applications.
A memristor's resistance evolves according to the total electric charge $q(t)$ that has flowed through it, enabling it to encode memory of past activity. Its behavior can be described by a generalized Ohm’s law:
\begin{equation}
    \label{eq:gen_ohm}
    V(t) = M\left( q(t) \right) I(t),
\end{equation}
where $M(q)$ is the memristance, a charge-dependent resistance function.

\begin{figure}[t]
    \centering
    \captionsetup{justification=raggedright}
    \includegraphics[width=\linewidth]{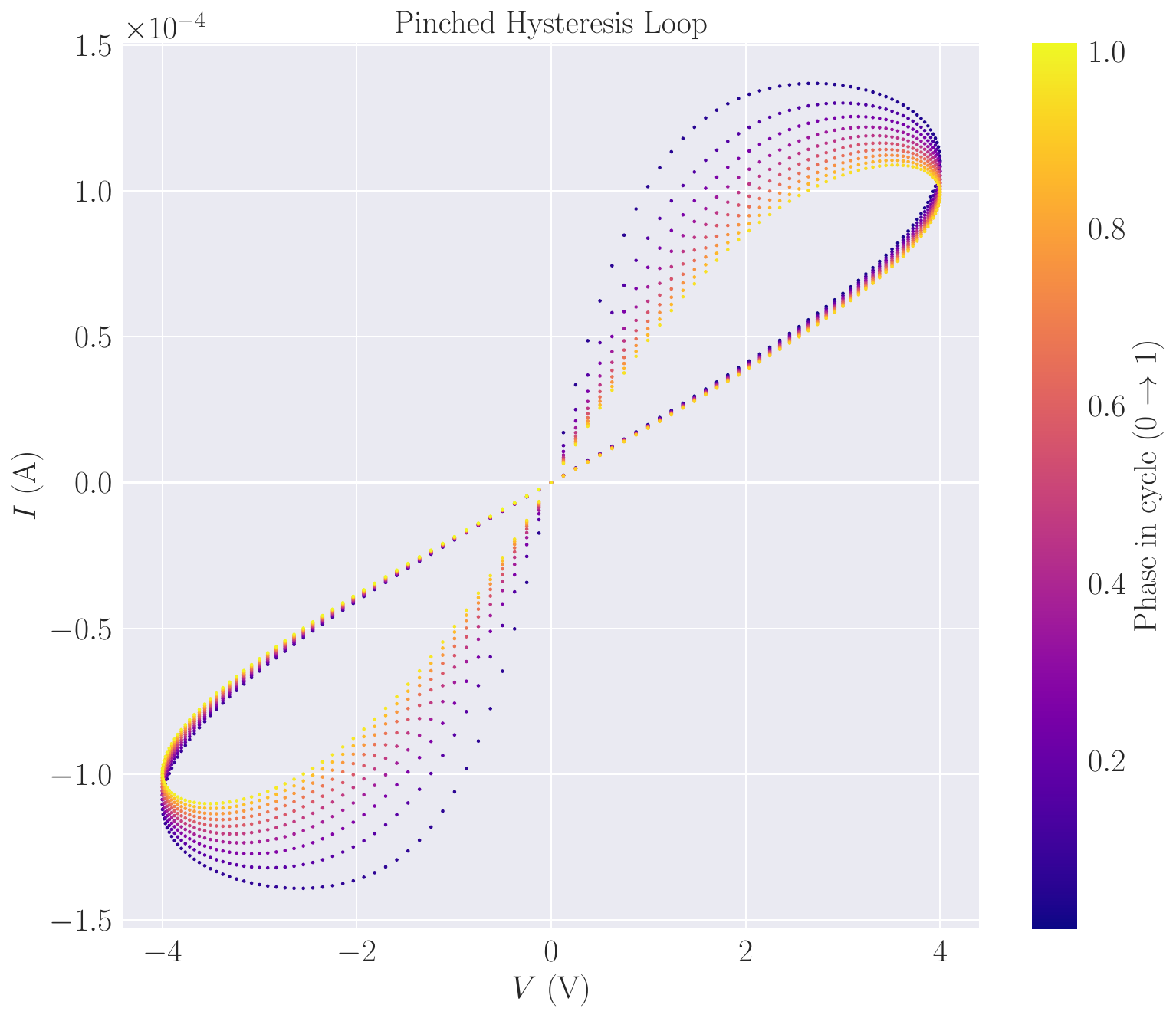}
    \caption{A pinched hysteresis loop in the $I$-$V$ plane of a memristor driven by a periodic input current $I(t) = I_0 \sin(\omega t)$. The parameters are $D=1\times10^{-9}$, $w=0.5\times10^{-9}$, $R_\mathrm{on}=1\times10^3$, $R_\mathrm{off}=1\times10^5$, $V_0=4.0$, $\omega=2\pi\times10^3$, over $2\,000$ time-steps.}
    \label{fig:mem_iv_hyst}
\end{figure}

When driven by a periodic signal, the current-voltage $I$-$V$ curve of a memristor displays a pinched hysteresis loop, which is a key experimental signature of memristive behavior\cite{chua_if_2014}, as shown in \Cref{fig:mem_iv_hyst}.

In this image, it can be seen that there is a clear figure-eight shape that is characteristic of a symmetric memristor, which is evolving in a clockwise loop starting from the origin. As it evolves through the phase of a full cycle of $0.02\,\si{\second}$, as shown by the color evolving from blue to yellow, the area covered by the loops begins to shrink as the memristance begins to collapse towards the diagonal, indicating the ``forgetting" of previous currents.

For a more realistic description, the Strukov model considers memristors implemented with thin-film titanium dioxide (TiO$_2$) devices \cite{strukov_missing_2008}. In this model, the device consists of two regions: a doped region with low resistance ($R_\mathrm{on}$) and an undoped region with high resistance ($R_\mathrm{off}$). The boundary between these regions is defined by the width $w(t)$ of the doped region, while the total thickness of the device is $D$.

The memristance as a function of charge can thus be expressed as:
\begin{equation}
    \label{eq:hp_mem}
    M(q(t)) = \left( R_\mathrm{on}\frac{w(t)}{D} + R_\mathrm{off} \left(1 - \frac{w(t)}{D}\right) \right).
\end{equation}
Here, the overall resistance is a weighted sum of the resistances of the two regions, determined by the proportion of the device occupied by each. This approach captures how the device's resistance changes dynamically as the boundary $w(t)$ shifts in response to the flow of charge, reflecting the underlying ionic drift mechanism in TiO$_2$ memristors.

\subsection{Memristive LIF Neuron}
\label{sec:Memristive_LIF_Neuron}

By replacing the leak resistor in the LIF circuit with a memristor, we obtain a memristive LIF neuron. This hybrid model introduces a key improvement: the leak conductance becomes state-dependent, allowing the neuron to adapt based on past input.

\begin{figure}[t]
    \centering
    \captionsetup{justification=raggedright}
    \includegraphics[width=\linewidth]{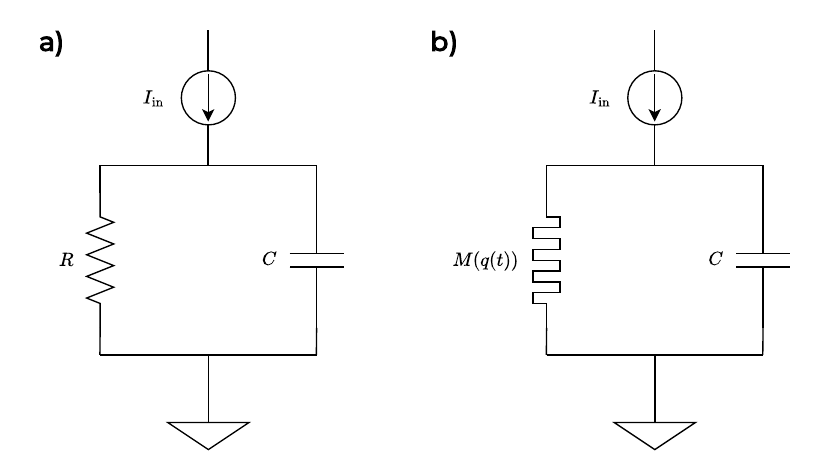}
    \caption{Leaky Integrate-and-Fire (LIF) neurons modeled as low-pass filter RC circuits. a) LIF circuit constructed with a resistor. b) LIF circuit constructed with a memristor in place of a resistor. The memristor is dependent on the history of the charge $q(t)$ that has passed through the circuit before.}
    \label{fig:lif_circuits}
\end{figure}

Substituting the resistor with a memristor $R\to M(q(t))$ in the LIF equation \Cref{eq:rc_diff} yields:
\begin{equation}
    \label{eq:mem_rc_diff}
    C_m \dv{V(t)}{t} = -\frac{V(t)}{M(q(t))} + I_\mathrm{in}(t),
\end{equation}
while the memristor state evolves through
\begin{equation}
    \label{eq:mem_rc_state}
    \dv{q}{t} = I(t) = \frac{V(t)}{M(q(t))}.
\end{equation}
This circuit forms the classical starting point for our quantization procedure. It captures both the spiking behavior of biological neurons and the memory-like adaptation of biological synapses and ion channels. The corresponding circuits are shown in \Cref{fig:lif_circuits}. 

\section{Quantization of the Circuit}
\label{sec:Quantization_of_the_Circuit}

To quantize the memristive LIF neuron, we address a key challenge: resistive elements such as memristors are inherently dissipative and thus incompatible with the unitary, reversible evolution of quantum systems. Although various quantum memristor models exist, here we pursue a physically grounded quantization within the LIF framework by modeling the memristive leak as an effective environment.

\subsection{Memristor as a Transmission Line Bath}
\label{sec:Memristor_Transmission_Line_Bath}

We replace the dissipative memristor with a semi-infinite, lossless transmission line (TL) composed of coupled \( LC \) oscillators, following the approach of the Caldeira-Leggett model~\cite{vool_introduction_2017}. This is done in line with the approach used in the counterpart framework used in the quantization of the Hodgkin-Huxley circuits \cite{gonzalez-raya_quantized_2019,gonzalez-raya_quantized_2020}.

The TL provides a broadband, approximately Ohmic admittance at the port, so that energy injected by the node propagates away without reflection. Dissipation and associated quantum noise then appear only in the reduced dynamics after tracing out the line, while the composite ``system + TL" remains lossless and amenable to canonical quantization.

The line acts as an absorbing bath, with its characteristic impedance \( Z_0 = \sqrt{L_0 / C_0} \) encoding the effective resistance. By adiabatically modulating \( Z_0(t) \), we mimic the memristor’s history-dependent behavior, allowing us to recover classical memristive dynamics in the weak-coupling, low-frequency limit.

\begin{figure}[t]
    \centering
    \captionsetup{justification=raggedright}
    \includegraphics[width=\linewidth]{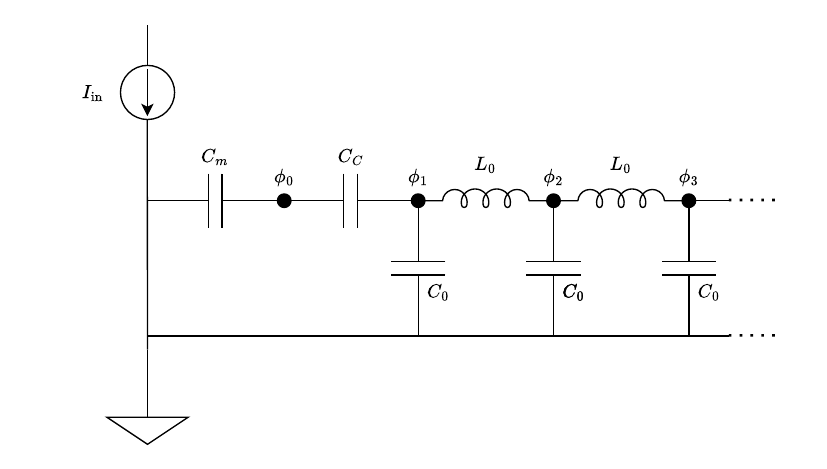}
    \caption{Memristive LIF circuit: the membrane node is driven by an AC current source \( I_{\mathrm{in}}(t) = I_0 \sin(\omega t) \), and weakly coupled via a capacitor \( C_C \) to a semi-infinite transmission line. Each node \( \phi_i \) in the TL has capacitance \( C_0 \) and inductance \( L_0 \), while \( C_m \) is the membrane capacitance.}
    \label{fig:tl_circuit}
\end{figure}

To couple the LIF node to this TL environment, we introduce a small coupling capacitor \( C_C \), as shown in \Cref{fig:tl_circuit}. This ensures that the interaction remains perturbative and does not dominate the system’s dynamics.

The system is described by the Lagrangian:
\begin{equation}
    \label{eq:lif_lagrangian}
    \begin{aligned}
        \mathcal{L} &= \frac{1}{2} C_m \dot{\phi}_0^2 + \frac{1}{2} C_C (\dot{\phi}_0 - \dot{\phi}_1)^2 \\
        &\quad + \sum_{i=1}^{\infty} \left[ \frac{1}{2} C_0 \dot{\phi}_i^2 - \frac{(\phi_{i+1} - \phi_i)^2}{2 L_0} \right],
    \end{aligned}
\end{equation}
where \( \phi_0 \) is the flux at the membrane node and \( \dot{\phi}_0 = V_0 \) is the membrane voltage.

Taking the continuum limit with spacing \( \Delta x \to 0 \), we define the flux field \( \phi(x,t) \), with:
\begin{equation}
    \label{eq:cont_lims}C_0 \Delta x \to \mathfrak{C}, \quad \frac{L_0}{\Delta x} \to \mathfrak{L},
\end{equation}
leading to the TL Lagrangian:
\begin{equation}
    \label{eq:tl_lagrangian}
    \mathcal{L}_\mathrm{TL} = \int_0^\infty \left[ \frac{1}{2} \mathfrak{C} (\partial_t \phi)^2 - \frac{1}{2\mathfrak{L}} (\partial_x \phi)^2 \right] \dd{x}.
\end{equation}
The wave equation for the TL field becomes:
\begin{equation}
    \label{eq:tl_wave}
    \mathfrak{C} \partial_t^2 \phi(x,t) = \frac{1}{\mathfrak{L}} \partial_x^2 \phi(x,t).
\end{equation}
At the membrane boundary \( x=0 \), the Euler-Lagrange equation yields:
\begin{equation}
    \label{eq:el_phi0}
    C_m \ddot{\phi}_0 - C_C \left[ \ddot{\phi}_0 - \partial_{tt} \phi(0,t) \right] - \frac{\phi_0}{L_m} = I_{\mathrm{in}}(t),
\end{equation}
where \( L_m \) is an effective inductance capturing the intrinsic membrane leak.

Under weak coupling \( C_C \ll C_m \), the correction to membrane inertia is negligible, giving:
\begin{equation}
    \label{eq:weak_coupling}
    C_m \ddot{\phi}_0 + \frac{1}{Z_0} \dot{\phi}_0 = I_{\mathrm{in}}(t),
\end{equation}
with \( Z_0 \equiv \sqrt{\mathfrak{L}/\mathfrak{C}} \). Recognizing \( \dot{\phi}_0 = V(t) \), we recover the classical memristive LIF equation.

\subsection{Quantization and Hamiltonian Formulation}
\label{sec:Quantization_Hamiltonian_Formulation}

We now construct a microscopic Hamiltonian by identifying the canonical momenta:
\begin{align}
    Q_0 &= \frac{\partial \mathcal{L}}{\partial \dot{\phi}_0} \approx C_m \dot{\phi}_0, \\
    \Pi(x) &= \frac{\delta \mathcal{L}}{\delta (\partial_t \phi)} = \mathfrak{C} \partial_t \phi(x,t).
\end{align}
The Hamiltonian is obtained via the Legendre transform:
\begin{equation}
    H = Q_0 \dot{\phi}_0 + \int \Pi(x) \partial_t \phi(x,t)\dd{x} - \mathcal{L}.
\end{equation}
This yields the total Hamiltonian:
\begin{equation}
    \label{eq:full_hamiltonian}
    \begin{aligned}
        H(t) =\;& \frac{Q_0^2}{2C_m} + \int_0^\infty \left[ \frac{\Pi^2}{2\mathfrak{C}} + \frac{(\partial_x \phi)^2}{2\mathfrak{L}} \right] \dd{x} \\
        & - C_C \frac{Q_0}{C_m} \Pi(0) - \phi_0 I_\mathrm{in}(t).
    \end{aligned}
\end{equation}
Promoting \( \phi_0 \) and \( Q_0 \) to operators with \( [\phi_0, Q_0] = i\hbar \), we obtain a fully quantum description of the circuit. The interaction term couples the membrane flux to the TL environment, enabling energy exchange and effective dissipation.

In the classical limit, Hamilton’s equations reproduce the expected dynamics:
\begin{align}
    \dot{\phi}_0 &= \frac{\partial H}{\partial Q_0} \approx \frac{Q_0}{C_m}, \\
    \dot{Q}_0 &= -\frac{\partial H}{\partial \phi_0} = I_{\mathrm{in}}(t) - \frac{V(t)}{Z_0}.
\end{align}
Combining these yields:
\begin{equation}
    C_m \dot{V}(t) + \frac{V(t)}{Z_0(t)} = I_\mathrm{in}(t),
\end{equation}
which identifies the effective memristive leak as \( Z_0(t) = M(q(t)) \).

We have derived a fully quantized model of the memristive LIF neuron by replacing the dissipative memristor with a semi-infinite transmission line bath. In the adiabatic, weak-coupling limit, this quantum model recovers the classical LIF neuron with a state-dependent leak conductance. This provides a rigorous microscopic foundation for future explorations of quantum neuromorphic systems.

\section{Numerical Simulations}
\label{sec:Numerical_Simulations}

With a now fully quantized description of the memristive LIF neuron circuit, we can verify the expected behaviors under a classical input drive, $I(t) = I_0\sin(\omega t)$, and investigate the validity of the outputs. This is done to obtain the characteristic feature of a memristor, as confirmation that the quantization procedure is indeed memristive. To do this, we will investigate the expectation values of the voltage and current of the quantized memristor under a period input drive, which should replicate the pinched hysteresis loop characteristic of memristors \cite{chua_if_2014}. Succeeding this we will investigate the full LIF neuron circuit, with appropriately included thresholding, firing, and reset mechanisms, to recover a consistent spiking neuron.

\subsection{Quantum Memristor}
\label{sec:Quantum_Memristor}

The characteristic feature of a memristor is the pinched hysteresis curve in the $I$-$V$ plane of the memristive response under a periodic input driving current. To verify that the quantum model is truly a memristor it needs to produce this characteristic behavior.

To produce the $I$-$V$ plane of the hysteresis plot, we can extract the expectation values of voltage, $V(t) = \expval{\dot{\hat{\phi}}_0}$, and current, $I(t) = \expval{\hat{\phi}}/Z_0(t)$, obtained when measuring the circuit. To simulate this, we integrate a time-dependent Gorini-Kossakowski-Sudarshan-Lindblad (GKSL) master equation \cite{breuer_theory_2010} while updating the memristance $Z_0(t) = M(q(t))$ at each of the $2\,000$ time steps of the density matrix evolution \cite{johansson_qutip_2012}.

\begin{figure}[t]
    \centering
    \captionsetup{justification=raggedright}
    \includegraphics[width=\linewidth]{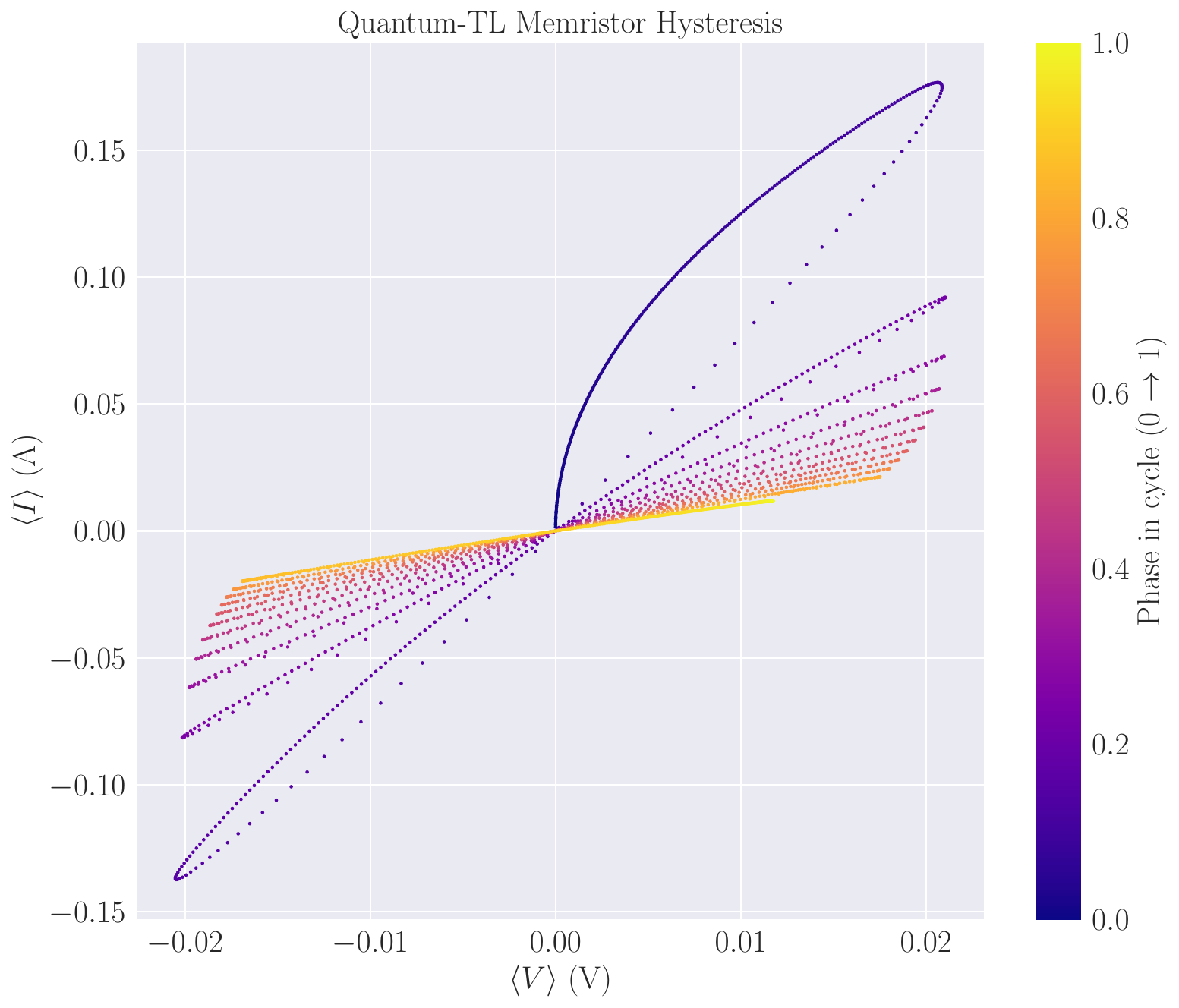}
    \caption{A pinched hysteresis loop in the $I$-$V$ plane of a quantum memristor driven by a periodic input current $I(t) = I_0 \sin(\omega t)$. The parameters are $C_m = 1.0$, $R_\mathrm{on}=1\times10^3$, $R_\mathrm{off}=1\times10^5$, $q_\mathrm{max}=1.0$, $I_0=1.0$, $\omega=\pi$, over $2\,000$ time-steps.}
    \label{fig:q_mem_iv_hyst}
\end{figure}

Looking at the single-mode oscillator of the membrane node, in terms of creation/annihilation operators $a^\dagger$/$a$, we have
\begin{align}
    \label{eq:single_mode_oscillators}
        \hat{\phi}_0 &= \sqrt{\frac{\hbar Z_0}{2}}(a + a^\dagger), \\
        \hat{Q}_0 &= \sqrt{\frac{\hbar}{2 Z_0}}(a^\dagger - a).
\end{align}
The reduced Hamiltonian for this case is then
\begin{equation}
    \label{eq:single_mode_hamiltonian}
        H(t) = \hbar \omega_0 \left( a^\dagger a + \frac{1}{2} \right) -\hat{\phi}_0 I_0 \sin(\omega t).
\end{equation}
We express the memristive leak in terms of a Lindbladian dissipator, in terms of annihilation operator $a$, as
\begin{equation}
    \label{eq:lindblad_dissipator}
    \mathcal{D}[a]\rho = a\rho a^\dagger - \frac{1}{2}\left\{ a^\dagger a, \rho \right\}.
\end{equation}
The associated decay rate
\begin{equation}
    \label{eq:gamma_q}
    \gamma(q) = \frac{1}{C_mM(q)}
\end{equation}
is tied directly to the instantaneous memristance $M(q)$, so that the strength of the dissipative channel follows the memristive state without introducing spurious dynamics.

The GKSL master equation is then expressed as
\begin{equation}
    \label{eq:lindblad_master}
    \dot{\rho} = -\frac{i}{\hbar} \left[H(t), \rho\right] + \gamma(t)\mathcal{D}[a]\rho,
\end{equation}
where the collapse factor for the master equation is $\gamma(t) = 1/(Z_0(t)C_m)$.

\begin{figure*}[t]
    \centering
    \captionsetup{justification=raggedright}
    \includegraphics[width=\textwidth]{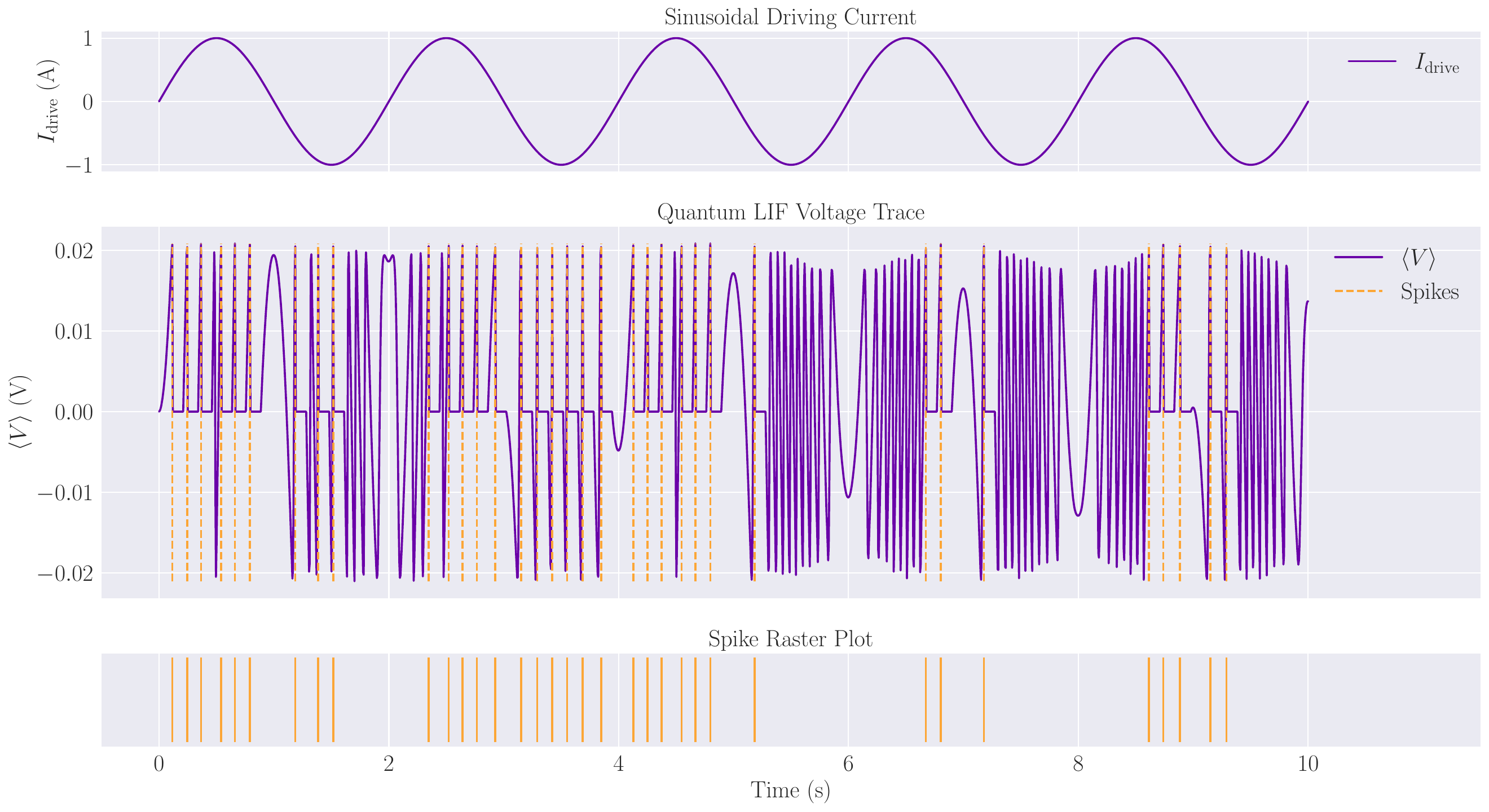}
    \caption{A quantum memristive leaky integrate-and-fire neuron being stimulated by a sinusoidal input current, producing output spikes based on incorporated thresholding and refractory mechanisms. The top pane is the input driving current stimulating the neuron circuit with an input current $I(t) = I_0 \sin(\omega t)$. The center pane is the voltage expectation value evolution under the driving current, with spikes indicated where the membrane threshold is crossed and the state is reset. The bottom pane is of the output spikes from the driven LIF neuron.}
    \label{fig:qlif_spikes}
\end{figure*}

The state update memristor law is adapted from the linear drift of \Cref{eq:hp_mem}, as we now have
\begin{align}
    \label{eq:mem_state_update}
        q(t) &= \int_{-\infty}^t \frac{\expval{\hat{\phi}_0}}{Z_0(\tau)}\dd{\tau}, \\
        Z_0(t) &= M(q(t)) \\
        M(q(t)) &= R_\mathrm{on}\left(\frac{q}{q_\mathrm{max}}\right) + R_\mathrm{off}\left(1-\frac{q}{q_\mathrm{max}}\right).
\end{align}
Numerically solving this master equation with the specified conditions, the pinched hysteresis loop in \Cref{fig:q_mem_iv_hyst} is obtained.

Comparing this to the classical hysteresis loop of \Cref{fig:mem_iv_hyst}, we notice that it has remarkably similar structure and behavior.
The differences in the quantum model having not as smooth and consistently drifting memristance is due to the approximations made in the adiabatic and weak coupling classical limits, as can be expected with any quantized model in this limit. Another reason is the quantum fluctuations present in the master equation formalism, with creation and annihilation operators being the quantum basis of the energy in the system. Additionally, one can see that the scales of the current and voltage are different, with the voltage in the quantum model being a factor of $10^{-3}$ smaller than the classical model, stemming from these losses. Nonetheless, the distinctly pinched hysteresis loop in the quantum $I$-$V$ plane is a clear verification of the model's validity as a true quantum memristor with non-zero memory. This model also scales directly with the ratios of the capacitance, voltage, and resistance being maintained. This stability is similar to the classical model; however, changes to the relative values of these parameters naturally create fluctuations in the model's behavior as a pure memristor. Contrasting with other formulations of quantum memristors, this model is based on a Markovian master equation, and stems from a canonical quantization perspective, rather than phenomenological modeling.

\subsection{Quantum Leaky Integrate-and-Fire Neuron}
\label{sec:Quantum_Leaky_Integrate-and-Fire_Neuron}

To extend the quantum memristor model to the full memristive quantum leaky integrate-and-fire (LIF) neuron, we must incorporate the key characteristics of LIF dynamics. Specifically, the model must not only integrate the input driving current but also implement a thresholding mechanism that triggers an output spike and resets the membrane potential. In our approach, the thresholding is monitored classically via the expectation value of the voltage operator. In the numerical simulation, this is done through interrupting the solver and replacing the state $\rho \to \dyad{0}$, which is the vacuum state, and continuing.

The use of the vacuum state injection lends itself to simple experimental implementations; however, those are beyond the scope of this demonstration, which is theoretical and computational. The intervention of classical monitoring and state-reset is not ideal for the similarity of a natural neuron. However, it is still a viable method for using this as a computational device.

Another key feature of the LIF neuron is the refractory period --- the brief interval following an output spike during which the neuron is temporarily unresponsive to further stimulation. In simulations, this is implemented by pausing both the input drive and memristor update for a duration $\tau_\mathrm{ref}$.

The operation of the memristive quantum leaky integrate-and-fire (LIF) neuron is depicted in \Cref{fig:qlif_spikes}, where a sinusoidal input current drives the quantum circuit to generate output spikes via a thresholding and refractory mechanism, mirroring the behavior of a classical LIF neuron.

Depending on system parameters, the spiking neuron exhibits various behaviors. However, the example in \Cref{fig:qlif_spikes} is particularly illustrative. The figure demonstrates a clear integration of the membrane potential in response to positive current, with analogous inhibitory behavior for negative current, reflecting the excitatory and inhibitory dynamics expected from biologically inspired models \cite{callaway_feedforward_2004, eshraghian_training_2023}. Refractory periods are evident as intervals with no updates to the membrane potential or memristance.

A notable feature of this example is the asymmetry in the output spike train, arising from the interplay between the timing and amplitude of the driving current. The current's amplitude is sometimes insufficient to reach the threshold, allowing the frequency-dependent oscillations to modulate the membrane potential while memristive dissipation damps the amplitude. Upon reversal of the current's sign, the membrane potential again grows toward the threshold.

\begin{figure*}[t]
    \centering
    \captionsetup{justification=raggedright}
    \includegraphics[width=\textwidth]{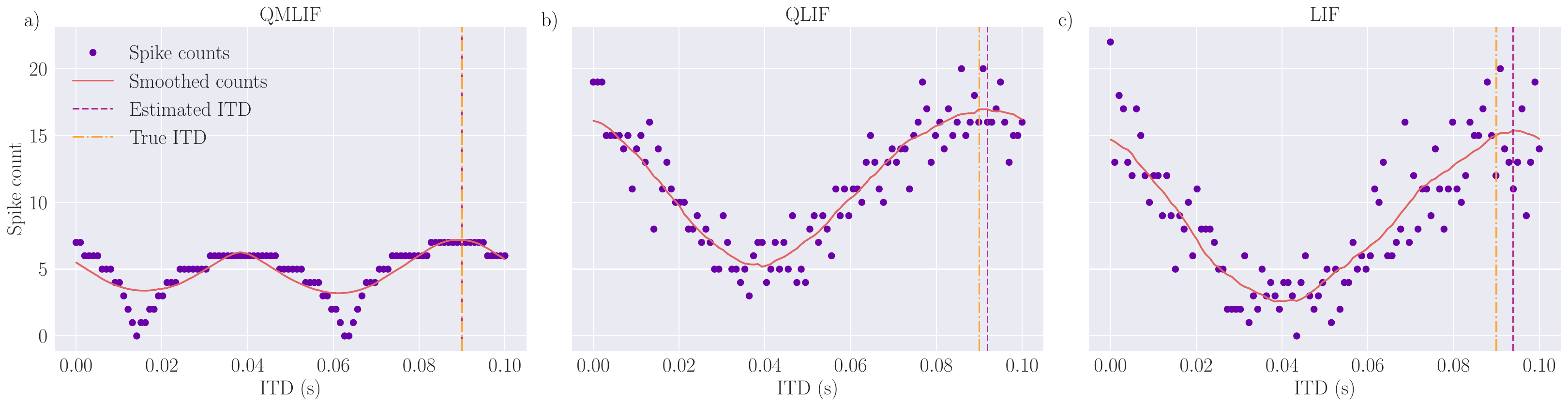}
    \caption{Spike-count-based decoding of interaural time differences (ITDs). Points indicate the raw spike counts per tested ITD; the dashed purple line marks the estimate, the dashed orange line the ground truth, and the red line across the points is a Savitzky-Golay filter smoother \cite{SavitzkyGolay1964}. The models tested are as follows: \textbf{(a)} Quantized memristive LIF (this work). \textbf{(b)} Phenomenological QLIF \cite{brand_quantum_2024}. \textbf{(c)} Classical LIF.}
    \label{fig:smoothed_cdsls}
\end{figure*}

\section{Sound Localization Through Coincidence Detection}
\label{sec:Sound_Localisation_Through_Coincidence_Detection}

One of the most important features of nervous systems in living creatures is the ability to localize sound sources in the environment. This is particularly crucial for vertebrates, which rely on auditory cues for survival, communication, and navigation. The Jeffress model of binaural coincidence detection provides a simple but clear framework for understanding how the brain processes incoming sound signals to determine the azimuthal direction of sound sources \cite{jeffressPlaceTheorySound1948}.

\begin{figure*}[t]
    \centering
    \captionsetup{justification=raggedright}
    \includegraphics[width=\textwidth]{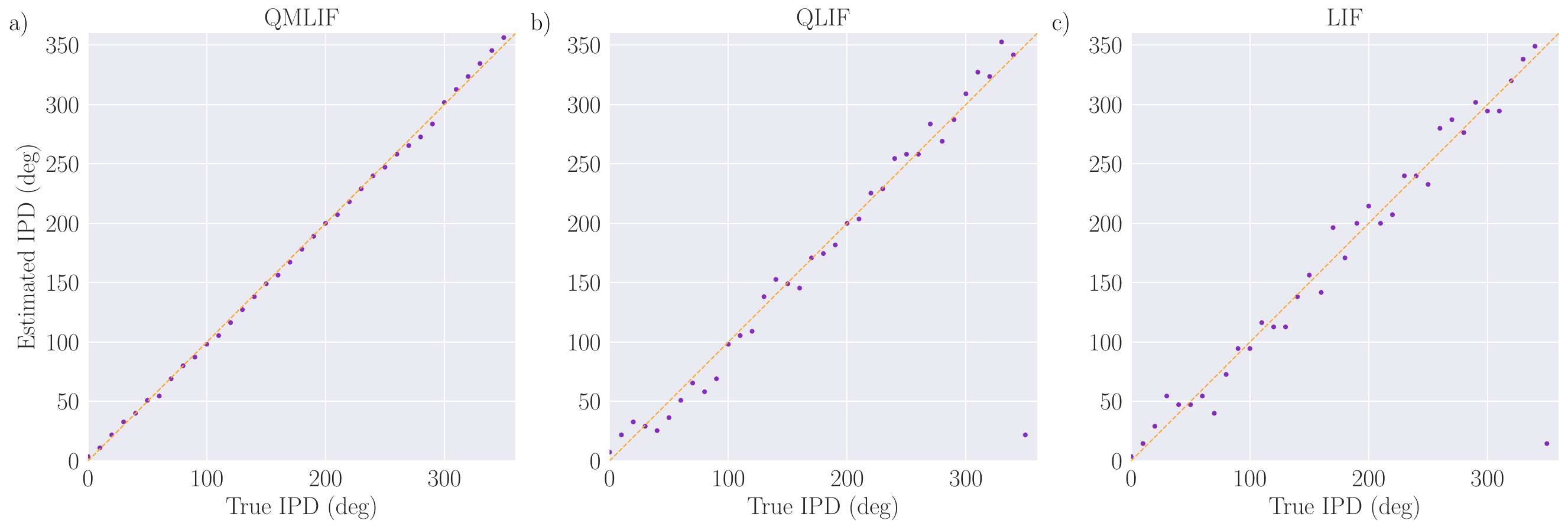}
    \caption{Interaural phase difference (IPD) decoding accuracy. Panels depict estimated versus true IPD; where the dashed diagonal is perfect performance. The models tested are as follows: \textbf{(a)} Quantized memristive LIF (this work). \textbf{(b)} Phenomenological QLIF \cite{brand_quantum_2024}. \textbf{(c)} Classical LIF.}
    \label{fig:slcd_err}
\end{figure*}

The first mechanical explanation for this is two parallel axonal delay lines (one from each ear) converging onto a linear array of coincidence detector neurons. Each detector neuron fires maximally when spikes from the left and right delay lines arrive simultaneously; as the conduction path length increases monotonically along the array, every physical interaural time difference (ITD) is mapped onto a unique neural locus. This model thus provides a place code for sound location, where the source angle is inferred from the identity, rather than the firing rate, of the most active neuron.

In the simulation of this model, we simplify the input auditory signals which are typically complex and noisy to a basic set of sinusoidal waveforms of the same frequency, shifted by a phase to represent the ITD. This allows for a controlled and repeatable set of inputs to the model, which can be used to evaluate the performance of the neurons in the Jeffress model.

The neurons that are stimulated the most indicate the greatest point of coincidence of the two signals traversing the brain. The ITD can be recast as an interaural phase difference (IPD), which better describes the angles of incidence to allow for better analysis of sound localization.

For this demonstration, three LIF models were examined to compare their performance and behaviors: a classical LIF, a phenomenological quantum LIF proposed by Brand et al. \cite{brand_quantum_2024}, and the quantized LIF presented here.

It can be seen in \Cref{fig:smoothed_cdsls} how for each neuron along the string through the brain in the Jeffress model there is a stimulation response in terms of how many output spikes are fired. By sweeping across all neurons, the maximally spiking neuron is found to be correspondent to the angle of incidence where the sound waves are coincidentally detected. As can be seen in the plots, the performance of the quantized LIF outperforms the phenomenological QLIF counterpart as well as the classical LIF, demonstrating its ability for neuromorphic tasks.

The map of sound localization across all angles is shown in \Cref{fig:slcd_err}, where it can be seen that the predicted and true incident angles are recorded for all angles and ideally lie along the diagonal. This benchmark shows that the quantized LIF presented here performs almost perfectly on a task representative of neuromorphic applications. Analysis of this model's trainability and performance in more challenging tasks and algorithms will be necessary to understand its full capabilities, and as such is planned for future study.

\section{Conclusion}
\label{sec:Conclusion}

We have presented a fully microscopic, quantized‑circuit derivation of the leaky integrate‑and‑fire neuron with a memristive leak. By coupling an $LC$ membrane oscillator to a semi‑infinite transmission‑line whose characteristic impedance $Z_0$ tracks the device's charge history, we constructed the complete Lagrangian and Hamiltonian, performed a Legendre transform, and showed that Hamilton’s equations reduce in the weak‑coupling, adiabatic limit to \Cref{eq:mem_rc_diff}, recovering the classical LIF equation. This derivation provides a principled foundation for understanding how memristive plasticity and quantized bath modes shape spiking‑neuron dynamics. The incorporation of this in a quantized spiking neuron circuit with demonstrable spiking behavior provides a unified framework for building upon this model towards more elaborate networks of quantum neurons, and for including further quantum properties and a new path for investigating usable memory in quantum circuits. The toy model demonstration of the quantized memristive LIF neuron on a sound localization neuromorphic benchmark task showed that it can outperform both the phenomenological QLIF and classical LIF models. Although this is a relatively simple task and not indicative of its fullest potential, it is a significantly encouraging outcome demonstrating its usability in neuromorphic processing. This potential can be explored in many ways, most notably with physical implementation and connection in networks of these neurons for greater computational power. With such potential and performance, the quantized memristive LIF neuron serves as a foundational framework for further expansion of quantum neuromorphic computing.

\section*{Data Availability}
The data presented in this article was created using Python, particularly the QuTiP package. The code created to obtain these results is available on GitHub at \url{https://github.com/deanbrand/QMLIFNC}.

\section*{Acknowledgements}
This work is based on the research supported in part by the National Research Foundation of South Africa, Ref. PMDS22070532362. The funders played no role in study design, data collection, analysis and interpretation of data, or the writing of this manuscript.

\section*{Author Contributions}
D.B. and D.D. conceptualised the quantization procedure. D.B. designed and executed numerical simulations. D.B. and D.D. wrote the manuscript. D.B. created the figures and plots. F.P. and D.D. supervised the research and provided valuable insight in the mathematical process and the writing of the manuscript. All authors reviewed and discussed the analyses and results.

\section*{Competing Interests}
The authors declare no competing interests.

\bibliographystyle{apsrev4-2}

\bibliography{references}

@article{pfeiffer_quantum_2016,
	title = {Quantum memristors},
	volume = {6},
	doi = {10.1038/srep29507},
	journal = {Scientific Reports},
	author = {Pfeiffer, P. and Egusquiza, I. L. and DI Ventra, M. and Sanz, M. and Solano, E.},
	month = jul,
	year = {2016},
    pages = {29507},
}

@article{sanz_invited_2018,
	title = {Invited article: {Quantum} memristors in quantum photonics},
	volume = {3},
	doi = {10.1063/1.5036596},
	number = {8},
	journal = {APL Photonics},
	author = {Sanz, M. and Lamata, L. and Solano, E.},
	month = aug,
	year = {2018},
    pages = {080801},
}

@article{markovic_physics_2020,
	title = {Physics for neuromorphic computing},
	volume = {2},
	doi = {10.1038/s42254-020-0208-2},
	number = {9},
	journal = {Nature Reviews Physics},
	author = {Marković, Danijela and Mizrahi, Alice and Querlioz, Damien and Grollier, Julie},
	month = sep,
	year = {2020},
	pages = {499-510},
}

@article{markovic_quantum_2020,
	title = {Quantum neuromorphic computing},
	volume = {117},
	doi = {10.1063/5.0020014},
	number = {15},
	journal = {Applied Physics Letters},
	author = {Marković, Danijela and Grollier, Julie},
	month = oct,
	year = {2020},
    pages = {150501},
}

@article{ghosh_quantum_2021,
	title = {Quantum {Neuromorphic} {Computing} with {Reservoir} {Computing} {Networks}},
	volume = {4},
	doi = {10.1002/qute.202100053},
	number = {9},
	journal = {Advanced Quantum Technologies},
	author = {Ghosh, Sanjib and Nakajima, Kohei and {Kohei Nakajima} and {Kohei Nakajima} and Krisnanda, Tanjung and {Keisuke Fujii} and {Keisuke Fujii} and Fujii, Keisuke and Liew, Timothy Chi Hin},
	month = jul,
	year = {2021},
	doi = {10.1002/qute.202100053},
	pages = {2100053},
}

@article{guo_y_-m_quantum_2021,
	title = {Quantum memristors with quantum computers},
	volume = {18},
	doi = {10.1103/PhysRevApplied.18.024082},
	number = {2},
	journal = {Physical review applied},
	author = {{Guo, Y. -M.} and {Albarrán-Arriagada, F.} and {Alaeian, H.} and {Solano, E.} and {Barrios, G. Alvarado}},
	month = dec,
	year = {2021},
    pages = {024082},
}

@article{chua_if_2014,
	title = {If it’s pinched it’s a memristor},
	volume = {29},
	doi = {10.1088/0268-1242/29/10/104001},
	number = {10},
	journal = {Semiconductor Science and Technology},
	author = {Chua, Leon O.},
	month = sep,
	year = {2014},
	doi = {10.1088/0268-1242/29/10/104001},
	pages = {104001},
}

@article{chua_memristive_1976,
	title = {Memristive devices and systems},
	volume = {64},
	doi = {10.1109/proc.1976.10092},
	number = {2},
	journal = {Proceedings of the IEEE},
	author = {Chua, Leon O. and Kang, Sung-Mo},
	month = feb,
	year = {1976},
	doi = {10.1109/proc.1976.10092},
	pages = {209--223},
}

@article{katumba_neuromorphic_2018,
	title = {Neuromorphic {Computing} {Based} on {Silicon} {Photonics} and {Reservoir} {Computing}},
	volume = {24},
	issn = {1077-260X, 1558-4542},
	url = {https://ieeexplore.ieee.org/document/8331848/},
	doi = {10.1109/JSTQE.2018.2821843},
	number = {6},
	urldate = {2023-03-13},
	journal = {IEEE Journal of Selected Topics in Quantum Electronics},
	author = {Katumba, Andrew and Freiberger, Matthias and Laporte, Floris and Lugnan, Alessio and Sackesyn, Stijn and Ma, Chonghuai and Dambre, Joni and Bienstman, Peter},
	month = nov,
	year = {2018},
	pages = {1--10},
}

@article{pehle_neuromorphic_2022,
	title = {Neuromorphic quantum computing},
	volume = {106},
	issn = {2470-0045, 2470-0053},
	url = {https://link.aps.org/doi/10.1103/PhysRevE.106.045311},
	doi = {10.1103/PhysRevE.106.045311},
	language = {en},
	number = {4},
	urldate = {2023-03-13},
	journal = {Physical Review E},
	author = {Pehle, Christian and Wetterich, Christof},
	month = oct,
	year = {2022},
	pages = {045311},
}

@inproceedings{sato_study_2004,
	address = {Budapest, Hungary},
	title = {A study on neuromorphic quantum computation},
	volume = {4},
	isbn = {978-0-7803-8359-3},
	url = {http://ieeexplore.ieee.org/document/1381200/},
	doi = {10.1109/IJCNN.2004.1381200},
	urldate = {2023-03-13},
	booktitle = {2004 {IEEE} {International} {Joint} {Conference} on {Neural} {Networks} ({IEEE} {Cat}. {No}.{04CH37541})},
	publisher = {IEEE},
	author = {Sato, S. and Kinjo, M. and Takahashi, O. and Nakamiya, Y. and Nakajima, K.},
	year = {2004},
	pages = {3253--3256},
}

@article{ghosh_quantum_2019,
	title = {Quantum reservoir processing},
	volume = {5},
	doi = {10.1038/s41534-019-0149-8},
	number = {1},
	journal = {npj Quantum Information},
	author = {Ghosh, Sanjib and Opala, Andrzej and Matuszewski, Michał and Paterek, Tomasz and Liew, Timothy Chi Hin},
	year = {2019},
	doi = {10.1038/s41534-019-0149-8},
	pages = {1--6},
}

@article{chua_memristor-missing_1971,
	title = {Memristor-{The} missing circuit element},
	volume = {18},
	doi = {10.1109/tct.1971.1083337},
	number = {5},
	journal = {IEEE Transactions on Circuit Theory},
	author = {Chua, Leon O.},
	month = sep,
	year = {1971},
	doi = {10.1109/tct.1971.1083337},
	pages = {507--519},
}

@article{strukov_missing_2008,
	title = {The missing memristor found},
	volume = {453},
	doi = {10.1038/nature06932},
	number = {7191},
	journal = {Nature},
	author = {Strukov, Dmitri B. and Snider, Gregory S. and Stewart, Duncan and Williams, R. Stanley},
	month = may,
	year = {2008},
	doi = {10.1038/nature06932},
	pmid = {18451858},
	pages = {80--83},
}

@article{tacchino_artificial_2019,
	title = {An artificial neuron implemented on an actual quantum processor},
	volume = {5},
	issn = {2056-6387},
	url = {https://www.nature.com/articles/s41534-019-0140-4},
	doi = {10.1038/s41534-019-0140-4},
	language = {en},
	number = {1},
	urldate = {2023-03-15},
	journal = {npj Quantum Information},
	author = {Tacchino, Francesco and Macchiavello, Chiara and Gerace, Dario and Bajoni, Daniele},
	month = mar,
	year = {2019},
	pages = {26},
}

@article{kristensen_artificial_2021,
	title = {An artificial spiking quantum neuron},
	volume = {7},
	issn = {2056-6387},
	url = {https://www.nature.com/articles/s41534-021-00381-7},
	doi = {10.1038/s41534-021-00381-7},
	language = {en},
	number = {1},
	urldate = {2023-03-28},
	journal = {npj Quantum Information},
	author = {Kristensen, Lasse Bjørn and Degroote, Matthias and Wittek, Peter and Aspuru-Guzik, Alán and Zinner, Nikolaj T.},
	month = apr,
	year = {2021},
	pages = {59},
}

@article{yamazaki_spiking_2022,
	title = {Spiking neural networks and their applications: a review},
	volume = {12},
	issn = {2076-3425},
	shorttitle = {Spiking neural networks and their applications},
	url = {https://www.mdpi.com/2076-3425/12/7/863},
	doi = {10.3390/brainsci12070863},
	language = {en},
	number = {7},
	urldate = {2023-03-28},
	journal = {Brain Sciences},
	author = {Yamazaki, Kashu and Vo-Ho, Viet-Khoa and Bulsara, Darshan and Le, Ngan},
	month = jun,
	year = {2022},
	pages = {863},
}

@misc{cao_quantum_2017,
	title = {Quantum {Neuron}: an elementary building block for machine learning on quantum computers},
	shorttitle = {Quantum {Neuron}},
	url = {http://arxiv.org/abs/1711.11240},
	urldate = {2023-05-17},
	publisher = {arXiv},
	author = {Cao, Yudong and Guerreschi, Gian Giacomo and Aspuru-Guzik, Alán},
	month = nov,
	year = {2017},
	note = {arXiv:1711.11240 [quant-ph]},
}

@article{gonzalez-raya_quantized_2020,
	title = {Quantized {Three}-{Ion}-{Channel} {Neuron} {Model} for {Neural} {Action} {Potentials}},
	volume = {4},
	issn = {2521-327X},
	url = {https://quantum-journal.org/papers/q-2020-01-20-224/},
	doi = {10.22331/q-2020-01-20-224},
	language = {en},
	urldate = {2023-05-17},
	journal = {Quantum},
	author = {Gonzalez-Raya, Tasio and Solano, Enrique and Sanz, Mikel},
	month = jan,
	year = {2020},
	pages = {224},
}

@article{wang_supervised_2020,
	title = {Supervised learning in spiking neural networks: a review of algorithms and evaluations},
	volume = {125},
	issn = {08936080},
	shorttitle = {Supervised learning in spiking neural networks},
	url = {https://linkinghub.elsevier.com/retrieve/pii/S0893608020300563},
	doi = {10.1016/j.neunet.2020.02.011},
	language = {en},
	urldate = {2024-04-04},
	journal = {Neural Networks},
	author = {Wang, Xiangwen and Lin, Xianghong and Dang, Xiaochao},
	month = may,
	year = {2020},
	pages = {258--280},
}

@article{waldrop_chips_2016,
	title = {The chips are down for moore’s law},
	volume = {530},
	url = {https://doi.org/10.1038/530144a},
	doi = {10.1038/530144a},
	journal = {Nature},
	author = {Waldrop, M. M.},
	year = {2016},
    pages = {144-147}
}

@article{merolla_million_2014,
	title = {A million spiking-neuron integrated circuit with a scalable communication network and interface},
	volume = {345},
	url = {https://doi.org/10.1126/science.1254642},
	doi = {10.1126/science.1254642},
	journal = {Science},
	author = {Merolla, P. A.},
	year = {2014},
    pages = {668-673}
}

@article{davies_loihi_2018,
	title = {Loihi: a neuromorphic manycore processor with on-chip learning},
	volume = {38},
	url = {https://doi.org/10.1109/MM.2018.112130359},
	doi = {10.1109/MM.2018.112130359},
	journal = {IEEE Micro},
	author = {Davies, M.},
	year = {2018},
    pages={82-99},
}

@article{benjamin_neurogrid_2014,
	author={Benjamin, Ben Varkey and others},
  journal={Proceedings of the IEEE}, 
  title={Neurogrid: A Mixed-Analog-Digital Multichip System for Large-Scale Neural Simulations}, 
  year={2014},
  volume={102},
  number={5},
  pages={699-716},
  doi={10.1109/JPROC.2014.2313565}
}

@article{zou_breaking_2021,
	title = {Breaking the von neumann bottleneck: architecture-level processing-in-memory technology},
	volume = {64},
	issn = {1674-733X, 1869-1919},
	shorttitle = {Breaking the von neumann bottleneck},
	url = {https://link.springer.com/10.1007/s11432-020-3227-1},
	doi = {10.1007/s11432-020-3227-1},
	language = {en},
	number = {6},
	urldate = {2024-06-05},
	journal = {Science China Information Sciences},
	author = {Zou, Xingqi and Xu, Sheng and Chen, Xiaoming and Yan, Liang and Han, Yinhe},
	month = jun,
	year = {2021},
	pages = {160404},
}

@book{breuer_theory_2010,
	address = {Oxford},
	edition = {Repr},
	title = {The theory of open quantum systems},
	isbn = {978-0-19-921390-0},
	publisher = {Clarendon Press},
	author = {Breuer, Heinz-Peter and Petruccione, Francesco},
	year = {2010},
}

@article{hodgkin_quantitative_1952,
	title = {A quantitative description of membrane current and its application to conduction and excitation in nerve},
	volume = {117},
	copyright = {http://onlinelibrary.wiley.com/termsAndConditions\#vor},
	issn = {0022-3751, 1469-7793},
	url = {https://physoc.onlinelibrary.wiley.com/doi/10.1113/jphysiol.1952.sp004764},
	doi = {10.1113/jphysiol.1952.sp004764},
	language = {en},
	number = {4},
	urldate = {2024-06-05},
	journal = {The Journal of Physiology},
	author = {Hodgkin, A. L. and Huxley, A. F.},
	month = aug,
	year = {1952},
	pages = {500--544},
}

@article{brand_quantum_2024,
	title = {A quantum leaky integrate-and-fire spiking neuron and network},
	volume = {10},
	copyright = {Creative Commons Attribution-NonCommercial 4.0 International License},
	issn = {2056-6387},
	url = {https://www.nature.com/articles/s41534-024-00921-x},
	doi = {10.1038/s41534-024-00921-x},
	language = {en},
	number = {1},
	urldate = {2025-04-04},
	journal = {npj Quantum Information},
	author = {Brand, Dean and Petruccione, Francesco},
	month = dec,
	year = {2024},
	pages = {125},
}

@article{vool_introduction_2017,
	title = {Introduction to quantum electromagnetic circuits},
	volume = {45},
	copyright = {http://onlinelibrary.wiley.com/termsAndConditions\#am},
	issn = {0098-9886, 1097-007X},
	url = {https://onlinelibrary.wiley.com/doi/10.1002/cta.2359},
	doi = {10.1002/cta.2359},
	language = {en},
	number = {7},
	urldate = {2025-06-26},
	journal = {International Journal of Circuit Theory and Applications},
	author = {Vool, Uri and Devoret, Michel},
	month = jul,
	year = {2017},
	pages = {897--934},
}

@article{johansson_qutip_2012,
	title = {{QuTiP}: {An} open-source {Python} framework for the dynamics of open quantum systems},
	volume = {183},
	copyright = {https://www.elsevier.com/tdm/userlicense/1.0/},
	issn = {00104655},
	shorttitle = {{QuTiP}},
	url = {https://linkinghub.elsevier.com/retrieve/pii/S0010465512000835},
	doi = {10.1016/j.cpc.2012.02.021},
	language = {en},
	number = {8},
	urldate = {2025-06-26},
	journal = {Computer Physics Communications},
	author = {Johansson, J.R. and Nation, P.D. and Nori, Franco},
	month = aug,
	year = {2012},
	pages = {1760--1772},
}

@article{callaway_feedforward_2004,
	title = {Feedforward, feedback and inhibitory connections in primate visual cortex},
	volume = {17},
	copyright = {https://www.elsevier.com/tdm/userlicense/1.0/},
	issn = {08936080},
	url = {https://linkinghub.elsevier.com/retrieve/pii/S0893608004000887},
	doi = {10.1016/j.neunet.2004.04.004},
	language = {en},
	number = {5-6},
	urldate = {2025-06-26},
	journal = {Neural Networks},
	author = {Callaway, Edward M},
	month = jun,
	year = {2004},
	pages = {625--632},
}

@misc{eshraghian_training_2023,
	title = {Training {Spiking} {Neural} {Networks} {Using} {Lessons} {From} {Deep} {Learning}},
	url = {http://arxiv.org/abs/2109.12894},
	doi = {10.48550/arXiv.2109.12894},
	urldate = {2025-06-26},
	publisher = {arXiv},
	author = {Eshraghian, Jason K. and Ward, Max and Neftci, Emre and Wang, Xinxin and Lenz, Gregor and Dwivedi, Girish and Bennamoun, Mohammed and Jeong, Doo Seok and Lu, Wei D.},
	month = aug,
	year = {2023},
	note = {arXiv:2109.12894 [cs]},
}

@article{gonzalez-raya_quantized_2019,
	title = {Quantized {Single}-{Ion}-{Channel} {Hodgkin}-{Huxley} {Model} for {Quantum} {Neurons}},
	volume = {12},
	doi = {10.1103/physrevapplied.12.014037},
	number = {1},
	journal = {Physical review applied},
	author = {Gonzalez-Raya, Tasio and Cheng, Xiao-Hang and Egusquiza, Iñigo L. and Chen, Xin and Chen, Xi and {Xi Chen} and Sanz, Mikel and Solano, Enrique},
	year = {2019},
	doi = {10.1103/physrevapplied.12.014037},
	pages = {014037},
}

@article{jeffressPlaceTheorySound1948,
  title = {A Place Theory of Sound Localization.},
  author = {Jeffress, Lloyd A.},
  year = 1948,
  journal = {Journal of Comparative and Physiological Psychology},
  volume = {41},
  number = {1},
  pages = {35-39},
  issn = {0021-9940},
  doi = {10.1037/h0061495},
  urldate = {2025-10-20},
  langid = {english}
}

@article{SavitzkyGolay1964,
  title = {Smoothing and {{Differentiation}} of {{Data}} by {{Simplified Least Squares Procedures}}.},
  author = {Savitzky, {\relax Abraham}. and Golay, M. J. E.},
  year = 1964,
  month = jul,
  journal = {Analytical Chemistry},
  volume = {36},
  number = {8},
  pages = {1627-1639},
  issn = {0003-2700, 1520-6882},
  doi = {10.1021/ac60214a047},
  urldate = {2025-10-22},
  langid = {english}
}

\end{document}